\documentclass[prb,twocolumn,showpacs,preprintnumbers,amsmath,amssymb,superscriptaddress,floatfix]{revtex4}
\usepackage{epsfig}
\usepackage{dcolumn}
\usepackage{amsmath}
\usepackage{color}
\usepackage{bm}

\begin{document}

\title{Infrared properties of Mg$_{1-x}$Al$_x($B$_{1-y }$C$_{y}$)$_2$ single crystals
in the normal and superconducting state}


\author{D. Di Castro}
\email[Email: ]{dicastro@physik.unizh.ch}

 \affiliation{"Coherentia" CNR-INFM and Dipartimento di Fisica,
Universit\`a di Roma La Sapienza, Piazzale Aldo Moro 2, I-00185 Roma, Italy}

\author{M. Ortolani}

  \affiliation{"Coherentia" CNR-INFM and Dipartimento di Fisica,
Universit\`a di Roma La Sapienza, Piazzale Aldo Moro 2, I-00185 Roma, Italy}

 \affiliation {Berliner Elektronenspeicherring Gesellschaft f\"{u}r Synchrotronstrahlung
 m.b.H (BESSY), Albert-Einstein-Str.15, D-12489
Berlin, Germany}

\author{E. Cappelluti}

\affiliation{SMC, ``Istituto dei Sistemi Complessi'', CNR-INFM, v. dei Taurini 19, 00185 Roma, Italy and
Dipartimento di Fisica, Universit\`a di Roma La Sapienza, Piazzale Aldo Moro 2, I-00185 Roma, Italy}

\author{U. Schade}
 \affiliation {Berliner Elektronenspeicherring Gesellschaft f\"{u}r Synchrotronstrahlung
 m.b.H (BESSY), Albert-Einstein-Str.15, D-12489
Berlin, Germany}

\author{N. D. Zhigadlo}
\author{J. Karpinski}

 \affiliation {Solid State Physics Laboratory, ETH, CH-8093
Z\"urich, Switzerland}

\date{\today}

\begin{abstract}
The reflectivity $R (\omega)$ of $ab$-oriented Mg$_{1-x}$Al$_x$(B$_{1-y }$C$_y$)$_2$ single crystals has been
measured by means of infrared microspectroscopy for $1300<\omega<17000$ cm$^{-1}$. An increase with doping of
the scattering rates in the $\pi$ and $\sigma$ bands is observed, being more pronounced in the C doped crystals.
The $\sigma$-band plasma frequency also changes with doping due to the electron doping, while the $\pi$-band one
is almost unchanged. Moreover, a $\sigma\rightarrow\sigma$ interband excitation, predicted by theory, is
observed at $\omega_{IB} \simeq 0.47$ eV in the undoped sample, and shifts to lower energies with doping. By
performing theoretical calculation of the doping dependence $\omega_{IB}$,  the experimental observations can be
explained with the increase with electron doping of the Fermi energy of the holes in the $\sigma$-band. On the
other hand, the $\sigma$ band density of states seems not to change substantially. This points towards a $T_c$
reduction driven mainly by disorder, at least for the doping level studied here. The superconducting state has
been also probed by infrared synchrotron radiation for $30<\omega<150$ cm$^{-1}$ in one pure and one C-doped
sample. In the undoped sample ($T_c$ = 38.5 K) a signature of the $\pi$-gap only is observed. At $y$ = 0.08
($T_c$ = 31.9 K), the presence of the contribution of the $\sigma$-gap indicates dirty-limit superconductivity
in both bands.
\end{abstract}
\maketitle

\section{Introduction}

Since the discovery of superconductivity in MgB$_2$ ($T_c
\simeq$40K)\cite{Nagamatsu} the electronic properties of this
system have been intensively studied from a theoretical and
experimental point of view. The band structure of MgB$_2$ is
characterized by two distinct electronic bands: the
 quasi two-dimensional  $\sigma$ band, formed by the
hybridized sp$_x$p$_y$ B orbitals and consisting of two holelike sheets, and  the three-dimensional $\pi$ band,
made of p$_z$ orbitals and consisting of  two  hole-like and one electron-like honeycombs.
\cite{Liu,Kortus,An,Uchiyama,Yelland,Eltsev} The disparity between $\pi$- and $\sigma$- band suppresses impurity
interband scattering giving rise to the most intriguing feature of the superconductor MgB$_2$, that is multigap
superconductivity. The $\sigma$ bands holes are strongly coupled to the in plane boron mode
E$_{2g},$\cite{An,Choi,Golubov,Kong} and originate a large superconducting gap ($\Delta_{\sigma}$), whereas a
small one opens on $\pi$-bands ($\Delta_{\pi}$). \cite{Szabo,Iavarone,Eskildsen,Wang,Bouquet,Tsuda,Quilty,Souma}
The electronic properties of MgB$_2$ can be modified by  chemical substitution.   The two elements which
substitute most readily in MgB$_2$ are Al for Mg and C for B
 (an exhaustive collection of literature data is
reported in Ref. \onlinecite{Kortus05}). Atomic substitutions change the electronic properties of MgB$_2$ both
by increasing the  impurity scattering (interband and intraband) and by changing the electron density. Indeed, C
and Al have one more electron than B and Mg respectively, and therefore it is expected that electrons are doped
into the system. The most evident effect of Al and C doping is the decrease of $T_c$ [see for example Refs.\
\onlinecite{Bianconi} and \onlinecite{Kazakov05} and references therein].

Two changes in the electronic properties with doping can be considered. The first change is the shift of the
Fermi level. Although electron doping is expected from the valence consideration, the experimentally observed
evidence is unclear. The change of the electron density was tentatively estimated by means of Hall effect
measurements, \cite{Masui} but a quantitative and selective analysis of Hall coefficient  is difficult in a
multi-band system. In MgB$_2$ the band filling due to electron doping should reduce the carrier density in the
hole like $\sigma$-band, but experimentally it is not yet clear if this effect leads to a decrease of the
density of states at the Fermi level $N_\sigma(0)$. The decrease of $N_\sigma(0)$, which enters electron-phonon
coupling constant $\lambda$, would lead to a reduction of $T_c$ and of the energy gaps $\Delta_\sigma,
\Delta_\pi$. The second change is the increase of carrier (inter-band and intra-band) scattering rate. The
latter effect was estimated by several experiments, as, for example, measurements of resistivity
 and upper critical field. \cite{Ohmichi,Ribeiro,Karpinski05,Kazakov05,Masui}
These effects are much stronger in C than in A substituted systems.
In particular, a quantitative estimate of the change of the intra-band scattering rate in the $\pi$ and $\sigma$
band, selectively, has been given by far infrared measurement of the reflectivity above and below $T_c$ in Al
doped and neutron irradiated polycrystalline samples.\cite{Ortolani} Evidence for the increase of the inter-band
scattering rate was given by measuring the merging of the two gap with C substitution. \cite{Gonnelli} Indeed,
while intra-band impurity scattering is expected to reduce $T_c$ by disorder-induced pair breaking, a small
inter-band impurity scattering may lead to a mix of $\sigma$ and $\pi$ Cooper pairs, averaging the order
parameters ($\Delta_{\sigma}\simeq$ 7meV and $\Delta_{\pi}\simeq2$meV) and reducing $T_c$ down to the isotropic
value. \cite{Golubov97}

Infrared (IR) spectroscopy, when performed on oriented samples, is a powerful tool to measure both the carrier
density and the scattering rate selectively in the two bands. Furthermore, IR spectroscopy is also a bulk probe
of the superconducting state, which can provide the gap values, as demonstrated recently in polycrystalline
MgB$_2$. \cite{Ortolani} Although more detailed information can be obtained in single crystals than in
polycrystalline samples, however the small size of  available MgB$_2$ single crystals (typically 0.3x0.3 mm$^2$)
has been  up to now a strong limit for the use of the IR spectroscopy. Indeed, this technique is challenging on
such samples, because of low signal intensity, further reduced by diffraction effects. For this reasons, until
now, an extensive IR spectroscopic work at low temperature has been carried out on polycrystals
\cite{Fudamoto,Kuzmenko01,Gorshunov,Ortolani} and oriented films \cite{Tu,Lobo} only. However, for a deeper
insight in the fundamental properties of MgB$_2$, oriented single crystal are needed, since in  films and
polycrystals both the intergrain effects and the increase in the scattering rate of the carriers are present and
may change the electronic properties qualitatively. The infrared response of  MgB$_2$ single crystals in the
superconducting state was first investigated by Perucchi \emph{et al.}\ \cite{Perucchi}  as a function of
temperature $T$ and magnetic field. To increase the signal in the far-infrared region ($25<\omega<50$
cm$^{-1}$), where conventional sources provide poor photon flux, Perucchi  \emph{et al.}\ built a mosaic of
several crystallites. Although clear effects of the superconducting transition could be observed in Ref.\
\onlinecite{Perucchi}, the reflectivity $R(\omega)$ was lower than 70\% for $50<\omega<6000$ cm$^{-1}$.
 More recently, Guritanu  \emph{et al.} \cite{Kuzmenko} investigated
the optical response of  single crystallites of MgB$_2$ in the mid-infrared and visible range at room $T$. They
found $R(\omega) > 90\%$ below 6000 cm$^{-1}$ with field parallel to the $ab$-plane. This value of $R(\omega)$
can
 be reconciliated with band calculation predictions \cite{belashenko,Ravindran}
 and dc-conductivity ($\sigma_{dc}=\rho^{-1}$) measurements,
 \cite{Kazakov05,Masui} indicating that the use of single crystallites seems
 to be preferred for an extensive study of the electronic
 properties by infrared spectroscopy.

In the present work we measure $R(\omega)$ in MgB$_2$ single crystal at room temperature, and investigate how it
is modified by C and Al doping. $R(\omega)$ in the pure sample is in agreement with
 the results of Ref. \onlinecite{Kuzmenko}.  Pure and doped crystals show a $R(\omega)$  metallic,
 with a pseudo plasma-edge at around 2 eV slightly decreasing on
doping. With increasing C or Al content, the expected increase of the scattering rates in the $\pi$
($\Gamma_\pi$) and $\sigma$ ($\Gamma_\sigma$) bands is observed. The plasma frequency of the $\sigma$ band is
affected by electron doping too, while that of the  $\pi$ band is almost constant. Furthermore, an absorption
band at $\sim 0.47$ eV is found in the pure sample, which becomes less evident and finally disappears as doping
proceeds. The latter feature, not observed in previous infrared experiments on pure MgB$_2$ single crystals,
\cite{Kuzmenko} is explained here in terms of $\sigma\to\sigma$ interband electronic transition by direct
comparison with band structure calculations. The calculated and observed redshift of the $\sigma$-interband
transition with C- and Al- doping allows us to provide an estimate of the corresponding Fermi level shift.

Since the $\sigma$ band in the pure compound fulfills the clean-limit condition ($\Gamma_{\sigma} \leq
2\Delta_{\sigma}$), \cite{Quilty} the superconducting transition is not expected to strongly modify the infrared
absorption around $\hbar\omega \simeq 2\Delta_{\sigma}$. This is not the case for the dirty-limit $\pi$-band.
Indeed previous infrared experiments in the superconducting state \cite{Perucchi, Tu} found an increase of
$R(\omega)$ around $\hbar\omega \simeq 2\Delta_{\pi}$ only. However, the increase of $\Gamma_{\sigma}$ is
expected to make the effect of the large gap $\Delta_{\sigma}$ observable in the far infrared spectrum, as
demonstrated in Ref. \onlinecite{Ortolani} on a neutron-irradiated polycrystalline sample. In the present work,
we show the different far-infrared response below $T_c$ of a pure and a C-doped single crystal, which can be
ascribed to the transition towards dirty superconductivity with C-doping in MgB$_2$.

\section{Experimental}

\begin{figure*}
        \includegraphics[width=18cm]{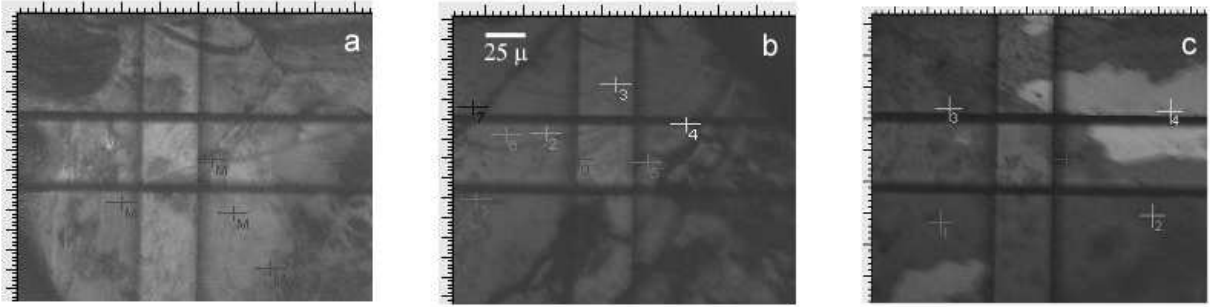}
    \caption{Images of three of the samples studied. The large tick intervals correspond to 25 $\mu$m.
        a. Pure MgB$_2$ b. Al-doped c. C-doped with  evaporated gold (clear region).
        The white crosses indicate the different
        points investigated to check possible inhomogeneity on the sample surface. The square in the center of
        the picture is the microscope clear aperture.}
    \label{foto}
\end{figure*}

We selected a series of C-doped and a series of Al-doped crystallites (see Table \ref{Tc}). The $ab$-oriented
surface is not larger than 300x300 $\mu$m in doped crystals, and it often presents cracks and holes when
inspected at the optical microscope (see Fig.\ref{foto}), thus making a normal-incidence reflectivity
measurement very challenging. However, we have selected clean surfaces for reflectivity measurements in the
mid-infrared range by means of an infrared microscope, where the beam coming from the interferometer is focused
in a micrometric spot-size by means of a beam condenser. However, infrared microscopy is limited on the
low-frequency side around 200 cm$^{-1}$ (radiation wavelength $\lambda$ of 50 $\mu$m) by diffraction effects.
The study of the superconducting state requires photon energies of the order of the gap or smaller, i.e. $\omega
\alt 100$ cm$^{-1}$ ($\alt$ 12 meV). Infrared Synchrotron Radiation (IRSR) with a conventional normal-incidence
reflectivity setup was used in the present work to provide a photon flux density high enough to get a signal in
the far-infrared from a 300x300 $\mu$m single crystal inside a cryostat.

 For the synthesis of Al and C
substituted MgB$_2$ crystals a high-pressure growth method  has been applied. \cite{Karpinski03} A detailed
description of the technique can be found in Refs. \onlinecite{Karpinski05} and \onlinecite{Kazakov05} for Al
and C doped system, respectively. The Al and C content of each  sample studied is listed in Table \ref{Tc}
together with superconducting $T_c$  determined by magnetization measurements.

\begin{table*}
\caption{List of single crystals Mg$_{1-x}$Al$_x($B$_{1-y }$C$_{y}$)$_2$ examined in the present work, with the
corresponding code name, C (y) and Al (x) content, and superconducting critical temperature $T_c$.}
\begin{ruledtabular}
\begin{tabular}{cccccc}
batch & code & x & y & $T_c$ (K)  \\
\hline
AN406/1 & P  & 0  &  0 & 38.5     \\
\hline
AN284/9 &C5.3 & 0  & 0.053 &  34.75   \\
AN286/6 &C8.3 & 0  & 0.083 &  31.9    \\
AN277/5 &C9.7 & 0  & 0.097 &  29.0    \\
AN258/2 &C11.4& 0  & 0.114 &  23.5    \\
\hline
AN262/19 &A7.2 & 0.072 & 0 &  33.05 \\
AN273/1  &A11.8 & 0.118 & 0 &  30.5 \label{Tc}
\end{tabular}
\end{ruledtabular}
\end{table*}

We used an Hyperion-2000 infrared microscope connected to a Bruker IF66 interferometer to select clear apertures
of 40x40 $\mu$m on the surface of each sample. Gold was evaporated on a portion of the sample surface to get an
in-situ reference signal (see Fig.\ \ref{foto}). The final $R(\omega)$ is obtained by multiplying for the
reflectivity of gold. A 40x40 $\mu$m clear aperture is large enough that there is no need to use the IRSR
source, so that a conventional thermal source and a tungsten filament lamp were employed. The reflectivity
$R(\omega$) in Fig.\ \ref{dataC} and \ref{dataAl} was measured at room temperature for $1300<\omega<17000$
cm$^{-1}$ with a 15x beam condenser. The low-frequency cut-on is due to the microscope detector, while the
high-frequency cut-off, close to the gold screened plasma frequency ($\sim$ 2 eV), is mainly due to the gold
film deposited on the microscope mirrors and to errors in the gold reflectivity correction procedure. The
response of MgB$_2$ single crystals in the superconducting state has been investigated in the present work by
normal-incidence reflectivity ratio measurements by FT-IR in the far-infrared range, where the photon energy is
of the order of the superconducting gap ($\hbar\omega \alt 10$ meV).

 The measurements at low-$T$ and
low-$\omega$ were performed at the infrared beamline IRIS at the BESSY-II storage ring in Berlin, Germany. The
beam collected at the end of a bending magnet from vertical angle of 40 mrad is sent to a Bruker IF66
interferometer equipped with a liquid-He cryostat and a normal-incidence reflectivity setup. A cut-on optical
filter at 100 cm$^{-1}$ was put in front of a 4.2 K bolometer. The intensity of the IRSR source is about one
order of magnitude larger than that of a Hg-arc lamp, depending on the current inside the storage ring (typical
values are 150 $\div$ 250 mA). However, the signal-to-noise ratio of the far-infrared data from a 300x300 $\mu$m
sample is strongly limited by several experimental problems, which we now discuss. Information on the gap value
of MgB$_2$ has been obtained from FT-IR measurements in far-infrared on pellets, \cite{Ortolani} films,
\cite{Tu} and single crystal mosaics. \cite{Perucchi} However, we decided to measure the intensity reflected by
one single crystal, not a mosaic or a film, in order to be able to compare the data with quantitative models of
the electrodynamic in-plane response, based on the Bardeen-Cooper-Schrieffer (BCS) theory, which includes the
plasma frequencies, the scattering rates and the superconducting gaps of both the $\pi$ and the $\sigma$ band
(see section IV). The disadvantage is that the available surface area for the reflection (300x300 $\mu$m$^2$) is
50 to 1000 times smaller than that, for example, of Refs. \onlinecite{Tu, Perucchi, Ortolani}. The loss of
signal intensity could in principle be recovered by the use of infrared synchrotron radiation (IRSR), which
provides a much higher brilliance if compared with blackbody sources. However, two main problems arise for
$\omega \to 0$: diffraction effects and increase in the IRSR beam size. In the far-infrared range ($20< \omega <
100$ cm$^{-1}$, $2.5 < E=\hbar\omega < 12.5$ meV) the radiation wavelength $\lambda$ is $0.1< \lambda < 0.5$ mm,
of the order of the sample size $d \simeq 0.3$ mm. Therefore, we are deeply into the diffraction regime, where a
large portion of the incoming radiation is not reflected at specular angle, but rather diffused in a much larger
solid angle. The effect of diffraction on the measured signal  is difficult to be exactly determined, as it
depends on the details of the experimental setup geometry. We have however estimated, through a series of test
measurements in our setup, that diffraction effects should decrease the signal intensity by one order of
magnitude in the entire far-infrared range. \cite{Jong} The second problem is that the IRSR beam size
approximately increases as $\lambda^{2/3}$ (Schwinger law). Since the spot size at the sample position in the
visible range is about 500 $\mu$m in diameter, the spot at 30 cm$^{-1}$ should be around 5 mm in diameter.
\cite{Ulli} As a consequence of this, only a minor part of the IRSR photons hit the sample.

To eliminate the diffraction effects, we eliminated any uncertainty on the optical alignment due to cryostat
strain or to IRSR variations by directly measuring the reflected intensity ratio (which in principle equals the
reflectivity ratio) $R_S/R_N =$ $R(\omega,T=4.2 K)/R(\omega, T=45 K)$ during several experimental runs without
heating above $T=50$ K. In this way, the diffraction pattern is washed out by dividing the spectra taken with
the same setup geometry. To avoid the contribution from the radiation not hitting the sample, we mounted the
crystals on the top of a cone-shaped sample holder. To check the contribution from diffuse radiation, we
performed a background intensity test by measuring the intensity reflected by the sample holder without the
sample. The background signal is smaller by a factor 20 and does not change for cryostat temperatures $T<50$ K,
and therefore could be easily subtracted. In conclusion, we could obtain reliable data for $30 < \omega < 100$
cm$^{-1}$ by averaging a total of 20000 interferometer scans, with a spectral resolution of 2 cm$^{-1}$ and an
uncertainty of $\pm 0.5\%$ on the reflectivity ratio. We repeated the procedure several times for one pure and
one C-doped sample ($y=0.083$).

\section{Normal state}

\begin{figure}
        \includegraphics{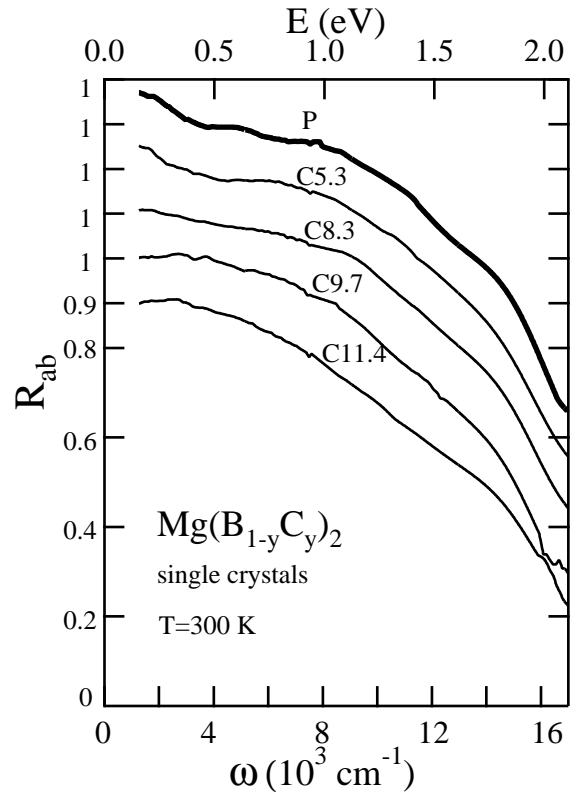}
    \caption{Reflectivity at 300K of C-doped crystals with field parallel to the {\em ab} plane.
    The curves are shifted for clarity.}
    \label{dataC}
\end{figure}

\begin{figure}
        \includegraphics{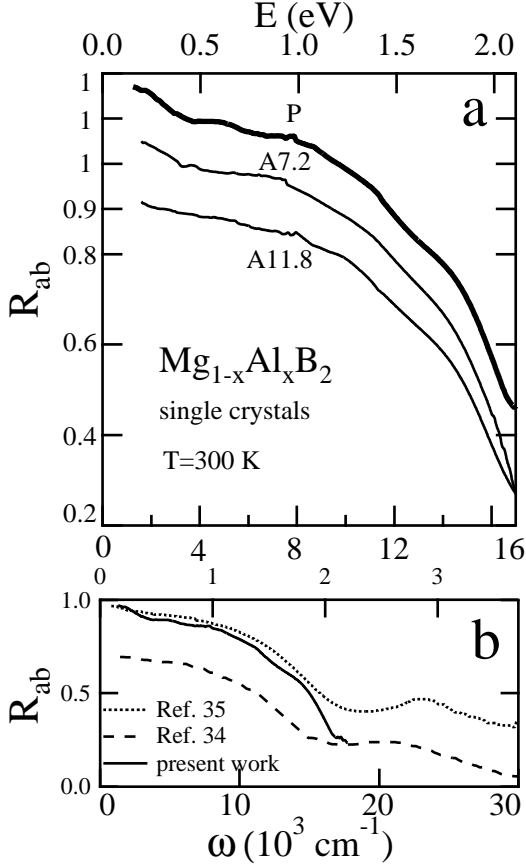}
    \caption{a. Reflectivity at 300 K of Al-doped crystals with field parallel
    to the {\em ab} plane. The curves are shifted for clarity.
    b. Comparison of $R(\omega)$ of pure MgB$_2$ single
    crystals from present work and from two previous works. \cite{Perucchi,Kuzmenko}}
    \label{dataAl}
\end{figure}

The reflectivity $R(\omega)$ at room temperature of the two single
 crystal series is shown in Fig.\ \ref{dataC} and \ref{dataAl} in
 the range of the present measurements ($0.2 < \omega < 2$ eV). For all crystals, $R(\omega)$
 is a typical free carrier response in a metal, with a pseudo plasma-edge (i.e. the frequency where $R(\omega)$
 almost reaches zero value) around 2 eV and almost
 independent on doping. The pseudo plasma-edge roughly corresponds to the
 screened plasma-frequency $\tilde{\omega}_p$, which is determined by  the free
 carrier density in both the $\sigma$ and $\pi$ bands by means of
the relation:

\begin{equation}
\tilde{\omega}_p^2 = \frac{\omega_{\pi}^2 + \omega_{\sigma}^2}{\epsilon_{\infty}} \label{omegaPl}
\end{equation}

\noindent where $\epsilon_{\infty}$ is the high frequency dielectric constant determined by the electronic
interband transitions and $\omega_{\sigma}, \omega_{\pi}$ are the plasma frequencies respectively for the
$\sigma$ and $\pi$ bands. We note that the level of the reflectivity in the mid-IR ($\omega < 1$ eV) decreases
with increasing  both Al and C doping, indicating the expected increase in the scattering  rates
$\Gamma_{\sigma}, \Gamma_{\pi}$ (please note that the curves  in Fig.\ \ref{dataC} and \ref{dataAl} are shifted
for clarity). In Fig.\ \ref{dataC} and \ref{dataAl} a kink at $\sim 0.4$ eV in $R(\omega)$ is evident in the
pure sample.
 This feature, not observed in previous infrared experiments on pure MgB$_2$
 single crystals,  becomes less prominent
 and finally disappears as doping proceeds.

Since there are several open questions about the quality of optical data
 on MgB$_2$ crystals, it is worth to compare in Fig.\ \ref{dataAl}-b our
 reflectivity spectrum on pure MgB$_2$ with those reported in
 Refs.\ \onlinecite{Perucchi} and \onlinecite{Kuzmenko}. The $\omega$-dependence of
 the free carrier contribution is remarkably similar in all
 three cases. A small discrepancy is seen below $\sim 0.5$ eV, where we observe a change of slope,
 not so clearly seen elsewhere. One may ask whether the present observation
  can be ascribed to surface contamination effects. Although the reason
  of this discrepancy is not clear at the moment, however, the reflectivity level
  in the present work is very close to that of Ref.\ \onlinecite{Kuzmenko}, where the surface contamination
  is minimal.
 Another discrepancy between our data and those of
  Ref.\ \onlinecite{Kuzmenko} is seen above 2 eV and probably due to an artifact of
  the gold reflectivity correction described in section I.

\begin{figure}
        \includegraphics[width=8cm]{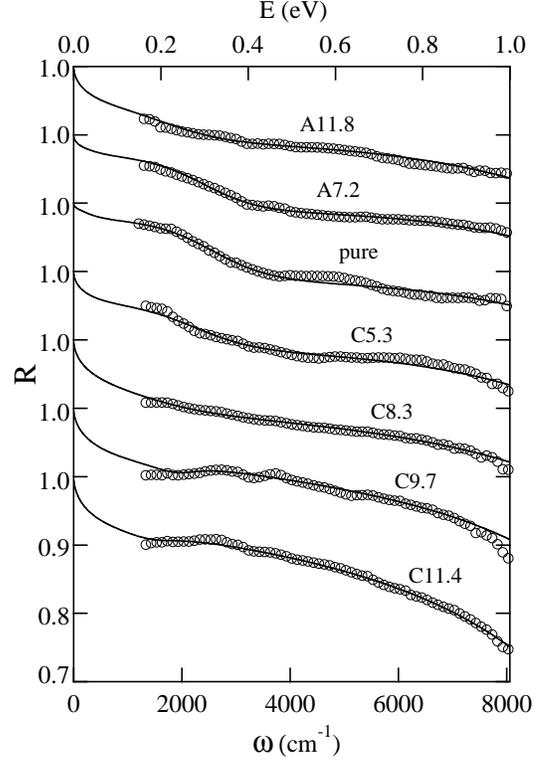}
    \caption{Reflectivity of pure, C-, and Al-doped MgB$_2$ crystals with field parallel to the {\em ab}
    plane in the low frequency range.
    The thin lines are the result of the fitting procedure described in the text.}
    \label{FigFit}
\end{figure}

The analysis of the optical spectra allows for a reliable determination of the plasma frequencies and the
scattering rates. Considering that it seems very hard to extract quantitative values for $\omega_{\pi}$ and
$\omega_{\sigma}$ from the measured screened plasma frequency, since the exact variation of $\epsilon_{\infty}$
with doping is unknown, we employed a Drude-Lorentz fitting procedure, which was shown to work well in the case
of pure MgB$_2$. \cite{Kuzmenko} The Drude model can be adapted to the case of MgB$_2$ by introducing two
separate Drude terms for the two bands, plus a suitable $\epsilon_{\infty}$ which we take as doping independent
and set to the value of 11.9 found by ellipsometric measurements in Ref.\ \onlinecite{Kuzmenko}. A Lorentzian
oscillator centered at $\omega_{MIR}$ = 0.47 eV was necessary to reproduce the change of slope seen in the
reflectivity. The final fitting formulas are therefore

\begin{eqnarray}
\tilde{\epsilon}(\omega)
&=& \epsilon_\infty -
\frac{\omega_{p,\sigma}^2}{\omega^2-i\omega\Gamma_\sigma}
-  \frac{\omega_{p,\pi}^2}{\omega^2-i\omega\Gamma_\pi}
\\ \nonumber
&& -\frac{S_{MIR}^2}{(\omega^2-\omega_{MIR}^2)-i\omega\Gamma_{MIR}},
\end{eqnarray}
\begin{equation}
R(\omega) = \left| \frac{1-\sqrt{\tilde{\epsilon}}(\omega)}{1+\sqrt{\tilde{\epsilon}}(\omega)}\right |,
\label{Rfit}
\end{equation}

\noindent where $S_{MIR}, \Gamma_{MIR}$ are the strength and the width of the Lorentzian oscillator
respectively. In Fig.\ \ref{FigFit} we show the result of the fitting procedure together with the reflectivity
data, which are well reproduced by the above formula. The discussion about the evolution with doping of the
resulting fitting parameters is detailed in the next section.

\begin{figure}
    \centering
        \includegraphics[width=8cm]{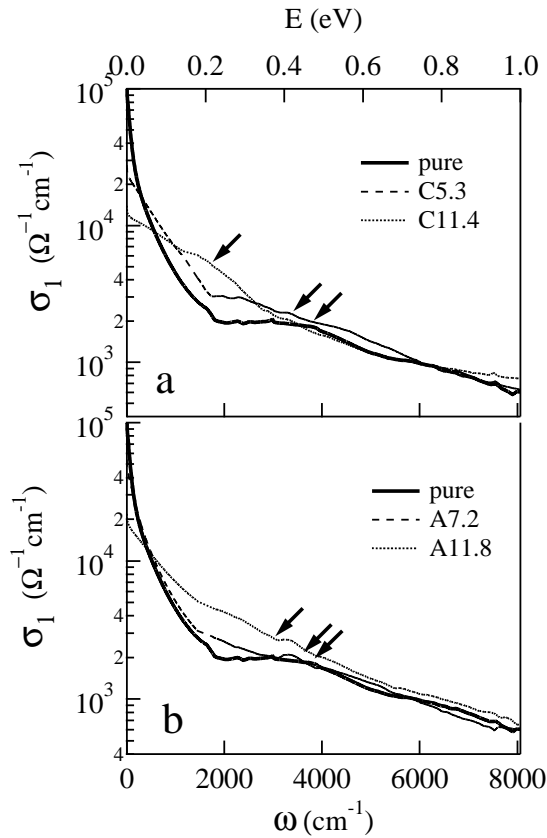}
    \caption{Optical conductivity of selected single crystals.
    The arrows indicate the value of $\omega_{MIR}$ as obtained from fitting the reflectivity data (see text).}
    \label{FigSigma}
\end{figure}

 The optical conductivity $\sigma(\omega)$ was obtained from
$R(\omega)$ through standard Kramers-Kronig transformations. The reflectivity was extrapolated to zero frequency
by using the fitting curves in Fig.\ \ref{FigFit} below 1300 cm$^{-1}$. To extrapolate at high frequencies we
used the data from Ref. \onlinecite{Kuzmenko} showed in Fig.\ \ref{dataAl}b. In Fig.\ \ref{FigSigma}, where
$\sigma(\omega)$ is plotted for representative doping values of C (panel a) and Al (panel b), a typical metallic
Drude-like behavior is clearly evident at low frequencies for all samples. The extrapolation of $\sigma(\omega)$
to zero frequency strongly decreases with doping in agreement with dc transport measurements on similar samples.
\cite{Kazakov05} At mid-IR frequencies, between 2000 and 5000 cm$^{-1}$, the deviation from a monotonous
behavior, more pronounced in the pure sample, indicates the presence of a broad absorption band superimposed to
the Drude peak. The black arrows indicate the frequency $\omega_{MIR}$ of the Lorentzian oscillator at 0.47 eV
introduced in the fit to the reflectivity data. The overall agreement between $\omega_{MIR}$ obtained from the
fit to $R(\omega)$ and the bump in the optical conductivity indicates that the change of slope in $R(\omega)$
corresponds to the mid-IR absorption band. The latter feature clearly shifts and broadens as the doping (Al and
C) increases. In the discussion section of this paper we will argue that this peak corresponds to an interband
electronic transition between two $\sigma$ bands.

\subsection{Discussion}

\begin{figure}
        \includegraphics[width=8cm]{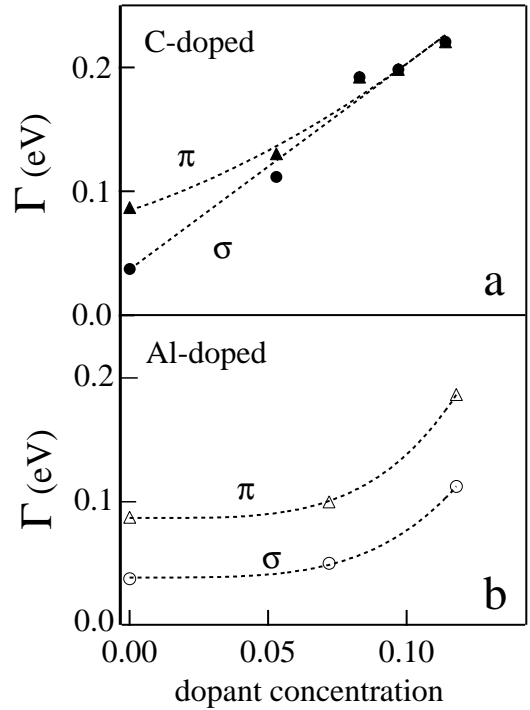}
    \caption{Scattering rates $\Gamma_{\sigma}, \Gamma_{\pi}$ as obtained from the
    fit of Eq.\ \ref{Rfit} to the reflectivity data, for C (panel a) and Al (panel b)
     doped crystals. The dotted lines are guides to the eyes.}
    \label{Gamma}
\end{figure}

The main effects of Al and C substitution are expected to be an increase of the scattering rate and a decrease
of the plasma frequencies in the $\sigma$ and $\pi$ bands. Looking at the Fig.\ref{Gamma} we notice a clear
evolution with doping of the scattering rate in the two bands, while the plasma frequency seems to vary only in
the $\sigma$-band. Let's first discuss the evolution of the scattering rates with doping.

 In Fig. \ref{Gamma} we
show  $\Gamma_{\sigma}$ and $\Gamma_{\pi}$ in the two series of Al and C doped samples as a function of dopant
concentration, that is the amount of substituted Mg or B atoms.

In the pure sample $\Gamma_{\sigma}=37$ meV and $\Gamma_{\pi}=87$ meV in good agreement with the values found in
Ref. \onlinecite{Kuzmenko}. This finding ($\Gamma_{\sigma} \ll\Gamma_{\pi}$) is expected, since the
$\sigma$-band is considered in a clean conduction regime, whereas the  $\pi$-band in a dirty one.
\cite{Quilty,Ortolani} With increasing Al and C doping the scattering rate increases in all samples of the two
series, but with a different trend. In the C doped samples $\Gamma_{\sigma}$ and $\Gamma_{\pi}$ both sharply
increase towards a common value at high doping level. This means that C substitution is much effective in
disordering  both the bands, but more strongly  the $\sigma$-band driving it towards a less clean conduction
regime. This is consistent with the strong increase of the upper critical field measured in C-doped single
crystals. \cite{Kazakov05,Masui} However, a different scenario could also be considered. In the present fitting
procedure, we can distinguish between the two bands only if they have a very different scattering rate.
Therefore, we cannot exclude that for high  C- doping levels, the conduction takes place in one band only. As
for the Al doped samples, the effect of doping on the intraband scattering rates is similar in the two bands:
they increase substantially only for $x > 0.07$, keeping $\Gamma_{\sigma} < \Gamma_{\pi}$. This result  suggests
that C substitution is more efficient than Al one in disordering both the bands.

\begin{figure}
    \centering
        \includegraphics[width=8cm]{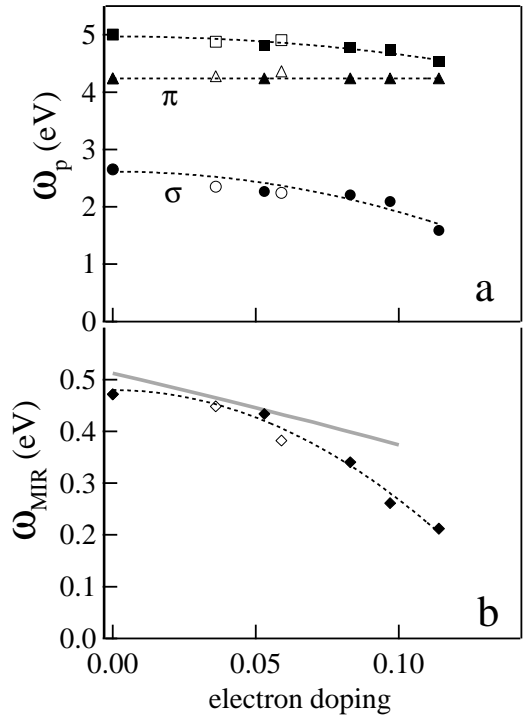}
    \caption{a. Plasma frequencies of the $\pi$ (triangles) and $\sigma$ (circles) bands and
    of the total plasma frequencies (squares) as a function of electron doping, as obtained from the fit of the
    Eq. \ref{Rfit} to the reflectivity data. b. Electron doping dependence of the absorption band given by
    the resonance frequency of the Lorentzian oscillator present in Eq. \ref{Rfit}. The grey thick line is the
    calculated doping dependence of the $\sigma\rightarrow\sigma$ interband transition (see text).
    Dashed line is a guide to the eyes. Empty symbols in both panels refer to Al-doped samples, filled symbols to the pure and
    C-doped samples.}
    \label{OmegaP}
\end{figure}

In addition to tune the magnitude of the disorder scattering processes, Al and C substitutions are expected to
introduce also additional charges in MgB$_2$ affecting thus the plasma frequency of the system. The measured
plasma frequencies $\omega_{p,\sigma}$, $\omega_{p,\pi}$, obtained from the fitting procedure of the
reflectivity data, as well as the ``total'' plasma frequency $\omega_p
=\sqrt{\omega_{p,\sigma}^2+\omega_{p,\pi}^2}$, are shown in Fig. \ref{OmegaP}a as a function of electron doping
for all the samples studied (full symbols refer to C doping and empty symbols to Al doped crystals).

For the undoped compound we find $\omega_{p,\sigma}=2.6$ eV, $\omega_{p,\pi}=4.2$ eV, yielding $\omega_p=5.0$
eV, much larger than  previously reported in optical experiments, apart from Ref. \onlinecite{Kuzmenko}. The
discrepancy between the present and the previous measured values of the plasma frequency can be ascribed to
different sample purity and air contamination, as also proposed in Ref. \onlinecite{Kuzmenko}. In the present
study, the samples are single crystals with minimal exposition to air, which explains the much larger value.
However, the present values, in particular $\omega_{p,\sigma}=2.6$ eV and $\omega_p=5.0$ eV, are still smaller
than those reported in Ref. \onlinecite{Kuzmenko} ($\omega_p \simeq$ 6.3 eV) and those predicted by the theory
($\omega_p \simeq$ 7 eV \cite{An,Kong,Liu}). The discrepancy between the theoretical and optically measured
values of the plasma frequencies has been addressed in Ref. \onlinecite{Kuzmenko}. We can note however that the
discrepancies between the present values and the ones reported in Ref. \onlinecite{Kuzmenko} can be explained by
the difference in the fitting procedure, since we added an absorption at 0.47 eV whose spectral weight is
comparable with that of the Drude terms. Different hypotheses can be invoked about the origin of this optical
structure. We cannot exclude for instance that a minimal exposition to air is responsible for shifting part of
the spectral weight from the Drude term to finite frequency absorption corresponding to charges localized by
defects in the crystal structure produced by air exposition. In this case, part of the spectral weight of the
band at 0.47 eV should be added to the computation of $\omega_p$ and a value around 6 eV could be easily
attained. However, as we will see later, we propose for the band at $\sim 0.47$ eV a different physical meaning
related to a predicted interband transition, as suggested by the doping dependence of its frequency
$\omega_{MIR}$.

As for the doping dependence of $\omega_{p,\nu}$ ($\nu = \sigma,\pi$), we note the $\pi$ band plasma frequency
does not change substantially within this electron doping range, whereas the $\sigma$ band presents a slight
decrease. The reduction of $\omega_{p,\sigma}$ can be easily explained in terms of reduction of the hole charge
density in the $\sigma$ bands which, in this doping regime, can be safely modelled as almost two-dimensional
parabolic bands. A simple calculation gives thus $\omega_{p,\sigma} \propto \sqrt{N_\sigma(0)E_F^\sigma
/m_\sigma^*}$, where $E_F^\sigma$ is the hole Fermi energy of the $\sigma$ bands which is roughly given by the
distance between the chemical potential and the top edge of the $\sigma$ bands. Theoretical calculations show
that the $\sigma$ band density of states $N_\sigma(0)$ is essentially doping independent in this range of
doping,\cite{An,Kortus} so that the decrease of $\omega_{p,\sigma}$ is mainly ruled by $E_F^\sigma$. According
to this analysis, we can thus estimate a reduction of $E_F^\sigma$ of a factor $\sim 0.36 $ from the pure to the
highest doped compound, which, using $E_F^\sigma \simeq 0.6$ eV for the undoped system,\cite{An,Kortus} gives
$E_F^\sigma(C11.4) \simeq 0.22$ eV for the C11.4 compounds, corresponding to s shift of the chemical potential
$\Delta\mu = \Delta E_F^{\sigma}\simeq 0.38$ eV. Note that, since the Fermi energy of the $\pi$ bands is much
larger than the $\sigma$ band one, $E_F^\pi \sim 5$ eV, the relative change in the $\pi$ band charge density is
much smaller and practically negligible, in agreement with Fig. \ref{OmegaP}a. We have to warn however that some
caution should be employed when extracting quantitative information about $\omega_{p,\sigma}$ and
$\omega_{p,\sigma}$ from the Drude-Lorentz fit in the high doping region. In this regime indeed the $\sigma$ and
$\pi$ band scattering rates $\Gamma_\sigma$, $\Gamma_\pi$ become very similar, making hard to separate in the
fit procedure the two distinct $\sigma$ and $\pi$ band contributions, whereas of course the total value of the
plasma frequency $\omega_{p}$ is not affected by this kind of problem.

Although a quantitative analysis of the $\omega_{p,\sigma}$, $\omega_{p,\pi}$ would requires some care in the
high doping range, a clearer evidence of the filling of the $\sigma$ bands as a function of the electron doping
may be given by the doping dependence of the resonance frequency of the Lorentzian oscillator present in Eq.
\ref{Rfit}. We have already mentioned above that a possible explanation for this structure could be related to
the interband
 $\sigma \to \sigma$ electronic transitions. An absorption band around 0.4-0.7 eV has been indeed
theoretically predicted in the pure sample, although not observed so far.\cite{Kuzmenko,belashenko}
However, there could be other different hypothesis to consider. One already  mentioned  possibility is the
presence of an absorption corresponding to charges localized by defects in the crystal structure produced by air
exposition. But the high values of both reflectivity data and plasma frequency seem to contradict this
hypothesis. Another possible interpretation of the MIR band could be given in terms of the frequency-dependent
electron-phonon scattering rate $\tau^{-1}(\omega)$. Indeed, if the conductivity is calculated from the
Eliashberg electron-phonon spectral function, \cite{Kuzmenko} then a deviation from the simple Drude model in
Eq.\ \ref{Rfit} is found in the MIR range, which can be interpreted as an increase of $\tau^{-1}(\omega)$.
However, due to the addition of dopant atoms, the scattering should be dominated by impurity scattering at all
$\omega$ (apart from that in the $\sigma$-band of the pure sample) and hence the conductivity could be as well
described by a frequency-independent $\tau^{-1} = \Gamma$ in Eq.\ \ref{Rfit} plus a MIR band, as for example in
the case of cuprates. \cite{Padilla} The presence of the MIR band with a similar intensity in all samples
suggests therefore that it might not be fully explained in terms of electron-phonon scattering.  A last
hypothesis could ascribe this MIR absorption band to a possible presence of MgO impurity on the sample surface.
The clear doping dependence  of the absorption frequency $\omega_{MIR}$ shown in Fig. \ref{OmegaP}b makes us
confident in excluding also this last possibility. Indeed, the softening of the $\omega_{MIR}$ with doping, as
shown in Fig. \ref{OmegaP}b, is expected since the distance between the two hole-like quasi-parabolic $\sigma$
bands decreases as electron charges are added to the compound.

To have a more quantitative insight, we compare these data with the simple model for the $\sigma$ bands
introduced in Ref. \onlinecite{An}, namely
\begin{equation}
\epsilon_{{\bf k},i}=\epsilon_0-(k_x^2+k_y^2)/m_i-2t_\perp \cos(k_zc),
\end{equation}
where $i=1,2$ is the index of the two $\sigma$ bands. Following Ref. \onlinecite{An} we take $\epsilon_0=0.6$
eV, $t_\perp=0.092$ eV, $m_1=0.20 m_e$, $m_2=0.53 m_e$. Within this model the $\sigma$ band IB absorption
results to have a energy window $0.26$ eV $< \omega < 1.29$ eV, and a Lorentzian fitting gives $\omega_{IB}
\simeq 0.51$ eV, in very good agreement with our data. We can also investigate the doping dependence assuming a
rigid band filling. It is easy to show that $\omega_{IB} \propto E_F^\sigma$. The results are shown in Fig.
\ref{OmegaP}b, also in an overall agreement with our data. The slight bend of our data with respect to the
almost linear behavior of the theoretical analysis can be probably ascribed to non-rigid band effects, which are
also known to be important in the doping range.\cite{Kortus05} Note that the reduction of the measured values of
$\omega_{IB}$, from the pure to the highest doped compound, is $\sim 55 \%$, also in a qualitative agreement
with the reduction of $E_F^\sigma$ obtained by $\omega_{p,\sigma}$. We conclude therefore that the assignment of
the $\sim 0.45$ eV absorption with the IB $\sigma$ transition is in good agreement with theoretical predictions,
thus giving  an experimental indication of an  upwards shift of the chemical potential with the substitution of
Al and C and of the corresponding electron doping of the $\sigma$ bands. As discussed before, since the $\sigma$
bands are almost two-dimensional and the corresponding DOS almost flat in this doping regime, the band filling
is however unlike to be the only source of the $T_c$ suppression, and we think that an important role is played
by the disorder as well.

\section{Superconducting state}

\begin{figure}
        \includegraphics[width=8cm]{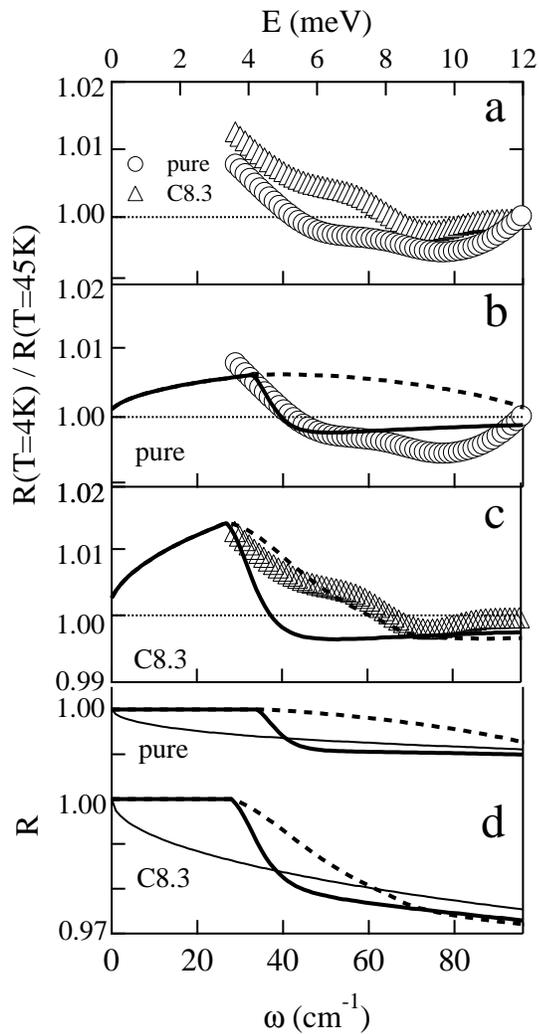}
    \caption{(a). Reflectivity change across the superconducting transition in one pure
    (circles) and one C-doped ($y$ = 0.083, triangles). The symbol size corresponds to
    the estimated standard deviation of the spectrum average. (b,c). The data from pure
    (b) and C-doped (c) crystals in a.\ are shown separately and compared to two BCS
    model curves (see text). Thick line: gap parameters in case 1  (parallel conduction),
    dashed line: gap parameters in case 2 (single effective gap). (d) The model reflectivity
    curves which generate the $R_S/R_N$ curves in (b,c). Thin line: normal state; dashed and thick
     lines represent the superconducting state in the case 1 and 2 respectively.}
    \label{rifSC}
\end{figure}

The reflectivity ratio ($R_S/R_N$) is reported for the pure and the C-doped sample in Fig.\ \ref{rifSC}-a. The
pure sample shows a decrease below 1 for $40 < \omega < 80$ cm$^{-1}$ and a sharper increase above 1 for $\omega
< 40$ cm$^{-1}$. The C-doped sample only shows an increase above 1 in $R_S/R_N$ for $\omega < 60$ cm$^{-1}$.
Above 80 cm$^{-1}$, there seems to be no effect of the superconducting transition on the $R_S/R_N$ within
uncertainties. Although the data quality does not allow an uncostrained best-fitting procedure, the fact that
both the present datasets are each taken on one oriented single crystal poses strong constraints. We are
therefore in the condition to compare the present data from MgB$_2$ samples with the prediction of different
BCS-based models for the infrared response of a two-gap superconductor. The BCS calculation employed here for
the complex conductivity $\tilde{\sigma}$ of one band in the superconducting state is that proposed by
Zimmermann \emph{et al.}. \cite{Zimmermann} Therein,  the normal state conductivity is a Drude term defined by
the plasma frequency $\omega_p$  and the scattering rate $\Gamma$, while the superconducting state conductivity
is computed trough the input of $T/T_c$ and the gap value $\Delta$.

However, in the present experiment the quantity which is probed is the reflectivity $R(\omega)$, rather than
$\tilde{\sigma}(\omega)$. In a s-wave BCS superconductor, for $\hbar\omega < 2\Delta$, the $R(\omega)$ in the
superconducting state is close to 100\%, because of total radiation screening from the supercurrent flowing in a
surface sheet, whose thickness is determined by the field penetration depth $\Lambda$. The normal state
reflectivity is then recovered even at $T \ll T_c$ for $\hbar\omega > 2\Delta$, since the radiation with energy
$\hbar\omega$ larger than that of the Cooper pairs $2\Delta$ cannot be screened. If one now turns to the two-gap
case, which is relevant for our work, one may ask whether the radiation would be screened or not for
$2\Delta_\pi < \hbar\omega < 2\Delta_\sigma$, since the photon energy is now larger than the $\pi$-band Cooper
pair energy, but smaller than the $\sigma$-band Cooper pair energy. Normal electrons, which are unpaired due to
radiation-induced Cooper-pair breaking in the $\pi$-band, will coexist with the surface supercurrent generated
by the pairs in the $\sigma$-band. The question whether the radiation with $2\Delta_\pi < \hbar\omega <
2\Delta_\sigma$ would be screened or not  has not yet been addressed theoretically to our knowledge. We
considered these two possibilities by using two different sets of parameters to calculate the conductivity in
the two bands (case 1 and case 2 described below). We then summed up the contribution from the two bands and we
determined the reflectivity and the $R_S/R_N$ from the complex conductivity with the standard Fresnel formulae.

In case 1 we assumed that the conductivity in the superconducting state is the parallel of two superconductors,
each one with a different $\omega_{p,\nu}, \Gamma_{\nu}$ and $\Delta_{\nu} (\nu=\sigma,\pi)$. This assumption,
which holds for the normal state, was already used to analyze the far-infrared data in the superconducting state
of MgB$_2$ pellets and films. \cite{Lobo,Ortolani} The superconducting state conductivity is  then given by
summing up the  contributions from the two bands, each weighted by $\omega_{p,\nu}^2$. The fact that $\Delta_\pi
\ll 3.52 k_BT_c$ must be explicitly taken into account in the finite temperature calculation. The use of the
parallel sum in the superconducting state corresponds to the assumption that there is no interaction between the
two superconducting fluids when they are exposed to an incoming electromagnetic wave. In case 2 we assume that
the unpaired $\pi$-electrons inhibit the radiation screening for frequencies $2\Delta_\pi < \hbar\omega <
2\Delta_\sigma$, thus allowing for the propagation of the radiation in the sample in a surface layer larger than
the theoretical value of $\Lambda_{\sigma}$. In this case, the reflectivity measurements are only sensitive to
the smaller gap $\Delta_\pi$. We  have therefore input in our calculations one single effective optical gap
$\Delta = \Delta_\pi$ for both bands in case 2. We note that the above discussion does not imply that there is a
single value for the gap  or for the penetration depth in the two-band superconductor, but rather one single
critical energy of the order of $2\Delta_\pi$ for the ac-field penetration depth.

We have generated two curves for each of the two samples, reported in Fig.\ \ref{rifSC}-b and -c, by using
different gap parameters for case 1 and case 2. The input parameters, reported in Table 2, are entirely
determined a priori, since the data quality does not allow a reliable fitting procedure. $\omega_{p,\pi}$ and
$\omega_{p,\sigma}$ were taken from the room temperature optical data reported in the present work on the same
samples, and assumed to be temperature independent. The scattering rates are mainly determined by impurity
scattering at low temperature. Therefore, we used the room temperature values for all the cases where impurity
scattering dominates the electron scattering also at room temperature, i.e. $\Gamma_\pi$ for the pure sample and
$\Gamma_\pi, \Gamma_\sigma$ for the C-doped sample. According to several authors, \cite{Kuzmenko,Kazakov05,
Masui} in the pure sample the electron scattering in the $\sigma$-band is dominated by phonons at room
temperature, so we cannot use the value for $\Gamma_\sigma$ from the present work. However, in Ref.
\onlinecite{Kuzmenko} an impurity scattering rate value of  12.4 meV was derived and found to be in agreement
with dc transport measurements; we have therefore used this value for the low-temperature $\Gamma_\sigma$ of the
pure sample. Concerning the gap values $\Delta_\pi$ and $\Delta_\sigma$, we have used the values from
photoemission measurements of Ref. \onlinecite{TsudaPES} for samples with similar doping level. $T/T_c$ was 0.1
for the pure sample and 0.13 for the C-doped sample.

\begin{table}
\caption{Input parameters used to generate the model curves in Fig.\ \ref{rifSC} and \ref{sigSC}. All data are
from the present work, apart $^a$ from Ref.\ \onlinecite{TsudaPES} and $^b$ from Ref.\ \onlinecite{Kuzmenko}.
All values are in meV, where not stated otherwise}
\begin{ruledtabular}
\begin{tabular}{cccccccc}
sample & $T_c$ (K) & $\Delta_\sigma$ & $\Delta_\pi$ & $\Gamma_\sigma$ & $\Gamma_\pi$ & $\omega_{p,\sigma}$ & $\omega_{p,\pi}$ \\
\hline
\\
$y=0$         & 38.5  &  6.4$^a$ & 2.1$^a$ & 12.4$^b$ & 100 & 2600 & 4200 \\
\\
$y=0.083$     & 31.9  &  4.2$^a$ & 1.7$^a$ & 180      & 180 & 2300 & 4200 \\
\end{tabular}
\end{ruledtabular}
\end{table}

The result of the analysis of Fig.\ \ref{rifSC} is the following: case 1, i.e. the parallel conductivity (dashed
line), can account for the data from the C-doped sample but not for those on the pure sample. On the other hand,
case 2, i.e. the single effective gap case (thick line) may explain the data from the pure sample. This is not
surprising, since \emph{all} the infrared reports of the superconducting gap value in undoped MgB$_2$ are in
agreement with the value for $\Delta_\pi$ from other techniques. \cite{Tu,Lobo,Perucchi} This low value of the
optical gap was often explained with the fact that the $\sigma$-band is a superconductor in the clean limit
($\Gamma \leq 2\Delta$) while the $\pi$-band is in the dirty-limit ($\Gamma \ > 2\Delta$). In a clean-limit
superconductor, the superconducting transition would not leave imprints in the electrodynamic response at the
gap energy, \cite{YouDoNotSee} and therefore infrared radiation would only probe the smaller $\pi$-gap. On the
other hand, the data on the C-doped sample, with both bands in the dirty limit, can be successfully reproduced
by a parallel sum of two dirty superconductors. As a consequence, we can confirm that the gap in the clean-limit
$\sigma$-band is not observed in the infrared experiments. On the other hand, even a fit curve which includes
the clean-limit condition can not reproduce the data for the pure sample. In the absence of more refined
theoretical models for the electrodynamic response of a two-gap superconductor, we speculate that the lack of
$\sigma$-gap signature in the reflectivity in the pure sample could be due to unpaired $\pi$-electrons
inhibiting the radiation screening for frequencies $2\Delta_\pi < \hbar\omega < 2\Delta_\sigma$. This is not the
case when the $\sigma$-band is also in the dirty limit, like in the C-doped sample, as shown by the higher
cutoff frequency in the $R_S/R_N$. We believe that the larger $\Lambda$, and hence the larger thickness of the
screening supercurrent sheet, may play a role in determining the higher cutoff frequency in the C-doped sample.
Note that, in spite of the higher cutoff frequency in the $R_S/R_N$, the absorptivity $1-R$ of the dirty C-doped
sample is larger than that of the pure sample, as expected (see Fig.\ \ref{rifSC}-d).

\begin{figure}
        \includegraphics[width=8cm]{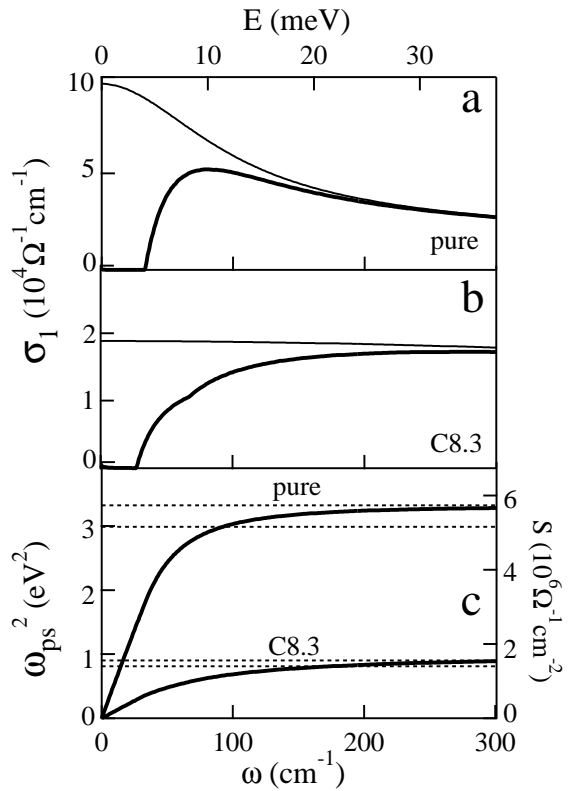}
    \caption{(a). Model optical conductivity of the pure sample in the normal (thin line)
    and superconducting (thick line) state as determined from case 2 gap values (see text).
    (b). The same as in (a) but for the C-doped sample, thick line obtained from case 1
    gap values. (c). The spectral weight difference between the normal and the superconducting
    state $S$ (right scale) for both the pure sample in (a) and the C-doped sample in (b).
    The thin dashed lines indicate the asyntotic value and 90\% of the asyntotic value
    for both samples. On the left scale the corresponding value of the squared
    superconducting plasma frequency is reported. $S$ of the pure sample shows a
    clean-limit behavior, since it reaches the 90\% line for $\omega \simeq 2\Delta_\sigma$,
    while the C-doped sample reaches the 90\% line at $\omega \simeq 6\Delta_\sigma$,
    hence showing a dirty-limit behavior.}
    \label{sigSC}
\end{figure}
The clean-limit nature of the $\sigma$-band in the pure sample can be made clear by looking, in Fig.\
\ref{sigSC}, at the model optical conductivity which was used to generate the $R_S/R_N$ curves in Fig.\
\ref{rifSC}. The normal-state conductivity in Fig.\ \ref{sigSC} is the sum of two purely Drude terms, and
therefore cannot be taken as a description of the real frequency dependence of the optical conductivity of
MgB$_2$ single cystals. However, since the model curves in Fig.\ \ref{sigSC} are obtained from a-priori values
of the input parameters and generate $R_S/R_N$ curves in good agreement with the experimental data, they can be
interestingly used to estimate for both samples the spectral weight loss below $T_c$. This quantity is given by

\begin{equation}
S(\omega) = \int_0^\omega \sigma_{1,n}(\omega ') - \sigma_{1,s}(\omega ')\ \rm{d}\omega '
\label{sweight}
\end{equation}

\noindent where $\sigma_{1,n},\sigma_{1,s}$ are the real part of the normal and superconducting state
conductivity respectively. The quantity $S(\infty)$, i.e. the value of the integral in Eq.\ \ref{sweight} for
$\omega \to \infty$, is proportional to the superfluid density, as the total spectral weight lost through the
transition is transferred to the zero-frequency delta-function representing the contribution of the
superconducting carriers to the optical conductivity (Ferrel-Glover-Thinkham sumrule).

 However, in our BCS approach, where the energy scale of the superconducting phenomenon is set by the larger
 energy gap $\Delta_\sigma$, 90\% of $S(\infty)$ can be recovered by extending the integral only up to
 few units of $\Delta_\sigma$, as shown by the dashed  lines in Fig.\ \ref{sigSC}-c. In the model
 for the pure sample (Fig.\ \ref{sigSC}-a) we find that 90\% of $S(\infty)$ is obtained by
 integrating up to 12 meV $\simeq 2\Delta_\sigma$ only. This correpsonds to the clean-limit case,
 since most of the normal-state spectral weight is located in the frequency region $\omega < 2\Delta$
 because of the small scattering rate $\Gamma_\sigma < 2\Delta$, which determines the width of the Drude peak.
 On the other hand, the model for the C-doped sample shows a broad Drude term and therefore a
 recovery of 90\% of $S(\infty)$ by integration up
 to 25 meV $\simeq 6\Delta_\sigma$  (being $\Delta_\sigma = 4.2$ meV in the C-doped sample)
 which is usually considered the upper limit of the integration in Eq.\ \ref{sweight} for
 dirty-limit BCS superconductors. \cite{Homes}

$S(\infty)$ can be more readily expressed in terms of a superconducting plasma frequency $\omega_{ps} =
\sqrt{(2/\pi)S(\infty)}$. $\omega_{ps}$ can be compared to the total plasma frequency $\omega_p$ derived from
the fit in Section 2 of the present work to give an estimate of the proximity of MgB$_2$ to the ideal
clean-limit condition, defined by the London model prediction $\omega_{ps}=\omega_p$ at $T \simeq 0$. We
obtained $\omega_{ps}= $ 1.8 and 0.9 eV for the pure and C-doped sample respectively (see Fig.\ \ref{sigSC}-c),
to be compared with $\omega_p$ = 5.0 and 4.7 eV respectively. We can conclude trough the analysis of the model
conductivity curves that the $R_S/R_N$ far-infrared data in Fig.\ \ref{rifSC} confirm the clean-limit scenario
for the pure sample and the the dirty-limit scenario for the C-doped sample with $y = 0.083$.

\section{Conclusions}
A systematic investigation of the in-plane optical properties of Mg$_{1-x}$Al$_x$(B$_{1-y }$C$_y$)$_2$ single
crystals has been curried out as a function of Carbon (x) and Aluminum (y) concentration by means of infrared
microspectroscopy for $1300<\omega<17000$ cm$^{-1}$.  For pure and doped crystals the reflectivity $R(\omega)$
is metallic, with a pseudo plasma-edge at around 2 eV slightly decreasing on doping.  Using a Drude-Lorentz fit
(see Eq. \ref{Rfit}) to the experimental $R(\omega)$, an increase of the scattering rates in the $\pi$
($\Gamma_\pi$) and $\sigma$ ($\Gamma_\sigma$) bands is found with increasing Al and C content. Whereas the
$\pi$-band plasma frequency results not to be affected by electron doping, the $\sigma$-band one shows a
substantial decrease. An absorption band at $\sim 0.47$ eV was found in the pure sample, which  becomes less
evident and finally disappears as doping increases. We performed band structure calculations which ascribe it to
a $\sigma\to\sigma$ interband electronic transition. The calculated and observed redshift of this transition
with C- and Al- doping allows us to provide an estimate of the corresponding Fermi level shift, which is in
qualitative agreement with the decrease of the $\sigma$-band plasma frequency. Since the $\sigma$ bands are
almost two-dimensional and the corresponding DOS almost flat in the doping region studied here, \cite{An,Kortus}
the band filling is  unlike to be the only source of the $T_c$ suppression, and we think that an important role
is played by the disorder as well.

The effect of doping in MgB$_2$ has been also probed in the superconducting state by using infrared synchrotron
radiation for $30<\omega<150$ cm$^{-1}$ in one pure and one C-doped single crystal. The  far-infrared response
below $T_c$ in the two samples is substantially  different, due to the different $\sigma$-band conduction
regime. Indeed,  in the undoped sample, with clean $\sigma$-band,  a signature of the $\pi$-gap only is
observed. In the C-doped one,  a contribution from the $\sigma$-gap to the reflectivity ratio appears,
indicating a transition towards dirty superconductivity with C-doping in MgB$_2$.

\section{Acknowledgement}
We are grateful to P. Dore, P. Calvani, P. Postorino, and H. Keller for illuminating and fruitful discussions.
We also acknowledge L. Baldassarre for help during measurements at the synchrotron BESSY in Berlin and M.
Cestelli Guidi, M. Piccinini and A. Nucara for help in preliminary measurements of the low temperature optical
response of single crystal.


\end{document}